\documentstyle[aps,epsfig,preprint,prb]{revtex}
\begin{document}
\draft

\title{Study of Harper's Equation
 for the 2-D Systems of Antiferromagnetically Correlated Electrons
 in an External Magnetic Field}
\author{Seung-Pyo Hong, Hyeonjin Doh, and Sung-Ho Suck Salk}
\address{Department of Physics, Pohang University of Science and Technology,
 Pohang 790-784, Korea}
\maketitle

\begin{abstract}
Considering interacting (antiferromagnetically correlated) electrons,
 we derive a generalized Harper's equation
for the square lattice of infinite size.
We obtain an analytic expression for the density of states 
 from the newly derived Harper's equation.
We present a predicted phase diagram of staggered magnetization
 in the plane of temperature vs doping rate 
 and discover a possibility of
 reentrant behavior of the staggered magnetization 
 even in the presence of applied magnetic field.
It is shown that
 below a critical electron correlation strength (Coulomb repulsion) 
 the staggered magnetization
 in the presence of magnetic field
 vanishes
 at an even denominator $q$ value
 but not at odd $q$
 of a given magnetic flux quantum per plaquette, $p/q$.
\end{abstract}

\vspace*{1cm}


\section{Introduction}
Since the discovery of high temperature superconductors
 and the related insulating materials,
 there has been a lot of interest in
 two-dimensional spin-$\frac{1}{2}$ magnetic
 or antiferromagnetically correlated electron systems.\cite{young,manousakis}
However not much attention has been paid to the magnetic
 properties of these systems
 coupled to an external magnetic field.\cite{bagehorn,held,scalapino}
The original Harper's equation \cite{hofstadter,macdonald,hasegawa}
 is concerned with the energy dispersion
 involving the systems of noninteracting electrons
 under the magnetic field.
Hence it is of great interest to study 
 how the systems of interacting
 (antiferromagnetically correlated) electrons behave
 under an external magnetic field.
We derive a generalized Harper's equation 
 which describes the dispersion of
 antiferromagnetically correlated  electrons
 under the applied magnetic field.
Earlier we paid attention to the dispersion of 
 the antiferromagnetically correlated electrons 
 only at half filling (and thus with no doping)
 and at zero temperature,
 by considering the square lattice of finite size.\cite{doh}
On the other hand,
 in the present study
 we derive a generalized Harper's equation
 for the square lattice of the infinite size,
 and examine the dispersion relation
 as a function of temperature and doping rate.
An analytic expression for the density of states is obtained
 from the generalized Harper's equation
 for the system of antiferromagnetically correlated electrons.
We present a phase diagram of staggered magnetization
 in the plane of temperature vs doping rate
 and find the hitherto-unnoticed reentrant behavior
 of the staggered magnetization
 even in the presence of external magnetic field.
Finally it is shown that
 below a critical electron correlation strength 
 the staggered magnetization disappears
 at an even denominator value of $q$ but not at odd $q$
 of a given magnetic flux quantum per plaquette, $p/q$.

\section{Generalized Harper's Equation and Density of States}
%for Antiferromagnetically Correlated Electrons
%in a Magnetic Field}
We write the Hubbard model Hamiltonian
 describing the two-dimensional system 
 of antiferromagnetically correlated electrons 
 under an external magnetic field,\cite{hasegawa}

\begin{equation} 
 H = -t \sum_{\langle ij\rangle\sigma}
  \left[
  \exp\left(-i\frac{2\pi}{\phi_0}\int_j^i {\bf A}\cdot d \bbox{\ell}\right)
  c_{i\sigma}^\dag c_{j\sigma}^{} 
  + \mbox{H.c.} \right]
  + U \sum_i n_{i\uparrow}^{} n_{i\downarrow}^{} 
  - \mu \sum_{i\sigma} c_{i\sigma}^\dag c_{i\sigma}^{}~~,
\end{equation}

\noindent
 where $t$ is the hopping integral;
 ${\bf A}$, the electromagnetic vector potential;
 $\phi_0=\frac{hc}{e}$, the elementary flux quantum; 
 $U$, the on-site Coulomb repulsion energy,
 and $\mu$, the chemical potential.
$\langle ij\rangle$ stands for summation 
 over nearest neighbor sites $i$ and $j$.
$c_{i\sigma}^\dag$ ($c_{i\sigma}^{}$)
 is the creation (annihilation) operator of an electron of
 spin $\sigma$ at site $i$,
 and $n_{i\uparrow}^{}$ ($n_{i\downarrow}^{}$), the number operator of
 an up-spin (down-spin) electron at site $i$.

The staggered magnetization (antiferromagnetic order)
 at site $i$ is written as
 $m_i =  e^{i{\bf Q}\cdot{\bf r}_i}
 \sum_\sigma \sigma \langle c_{i\sigma}^\dag c_{i\sigma}^{} \rangle$,
 where ${\bf Q}=(\pi,\pi)$
 and ${\bf r}_i=(i_x,i_y)$ with $i_x$ and $i_y$ being integers 
 with the lattice spacing of unity.
Introducing
 a uniform staggered magnetization $m$
 and a uniform doping rate $\delta$, i.e.,
\addtocounter{equation}{1}
\newcounter{uniform}
\setcounter{uniform}{\arabic{equation}}
$$
 m = \frac{1}{N} \sum_{i\sigma} e^{i{\bf Q}\cdot{\bf r}_i}
     \sigma \langle c_{i\sigma}^\dag c_{i\sigma}^{} \rangle ~~,
 \eqno{(\arabic{uniform}a)}
$$
$$
 \delta = 1 - \frac{1}{N} \sum_{i}
          \langle n_i^{} \rangle ~~,
 \eqno{(\arabic{uniform}b)}
$$
 with the number of lattice sites $N$,
 and using the Landau gauge ${\bf A}=B(0,x,0)$,
 we obtain the mean field (Hartree-Fock) Hamiltonian
 in the momentum space,
\begin{eqnarray}
 H &=& -t \sum_{{\bf k}\sigma}
  \left[2\cos k_x c_{{\bf k}\sigma}^\dag c_{{\bf k}\sigma}^{} 
  + e^{-ik_y} c_{{\bf k}-{\bf g},\sigma}^\dag c_{{\bf k}\sigma}^{} 
  + e^{ik_y} c_{{\bf k}+{\bf g},\sigma}^\dag c_{{\bf k}\sigma}^{}\right]
 \nonumber \\
 && -\frac{mU}{2}\sum_{{\bf k}\sigma} \sigma
  c_{{\bf k}+{\bf Q},\sigma}^\dag c_{{\bf k}\sigma}^{} 
  + [\frac{U}{2}(1-\delta)-\mu] \sum_{{\bf k}\sigma} 
  c_{{\bf k}\sigma}^\dag c_{{\bf k}\sigma}^{}~~, 
\label{hammag}
\end{eqnarray}
 where
 ${\bf g}\equiv \left(2\pi\frac{\phi}{\phi_0},0\right) = \left(2\pi\frac{p}{q},0\right)$
 with $\frac{p}{q}$, the number of flux quanta per plaquette.
The first bracketed term in (\ref{hammag}) represents hopping processes;
 the first term in the bracket 
 represents the nearest neighbor hopping in the $x$-direction
 and the last two terms in the bracket,
 the nearest neighbor hopping in the $y$-direction.
Because of the choice of the Landau gauge ${\bf A}=B(0,x,0)$,
 the electron acquires no phase when it hops in the $x$-direction,
 while it acquires a phase when it hops in the $y$-direction,
 and the electromagnetic vector potential ${\bf A}$ 
 shifts the wave vector of electron in the $k_x$-direction
 by $g \equiv |{\bf g}|=2\pi\frac{p}{q}$.
The second term results from the nature of
 the antiferromagnetic spin order of correlated electrons,
 which causes the wave vector to shift by ${\bf Q}$.
The last term represents energy shift by $\frac{U}{2}(1-\delta)$
 as a result of hole doping.

The Hamiltonian (\ref{hammag}) can be written as
$$
H = H_0 + H_1 ~,
$$
$$
 H_0 = [\frac{U}{2}(1-\delta)-\mu] \sum_{{\bf k}\sigma}
  c_{{\bf k}\sigma}^\dag c_{{\bf k}\sigma}^{}~~, 
$$
\addtocounter{equation}{1}
\newcounter{Hks}
\setcounter{Hks}{\arabic{equation}}
$$
 H_1 = \sum'_{{\bf k}\sigma}
 {\bf C}^\dag_{k\sigma} {\bf H}_{k\sigma} {\bf C}_{k\sigma} ~,
\eqno{(\arabic{equation}a)}
$$
where
$$
{\bf C}_{k\sigma} =
        \left[ \begin{array}{c}
         c_{{\bf k}+{\bf g},\sigma} \\
         \vdots \\[-2mm]
         c_{{\bf k}+(q-1){\bf g},\sigma} \\
         c_{{\bf k}\sigma} \\
         \hline
         c_{{\bf k}+{\bf g}+{\bf Q},\sigma} \\
         \vdots \\[-2mm]
         c_{{\bf k}+(q-1){\bf g}+{\bf Q},\sigma} \\
         c_{{\bf k}+{\bf Q},\sigma}
        \end{array} \right] ~,
\eqno{(\arabic{equation}b)}
$$
$$
  {\bf H}_{k\sigma} = \left[ \begin{array}{c|c} {\bf T}_k & {\bf V}_\sigma \\ \hline {\bf V}_\sigma & -{\bf T}_k \end{array} \right] ~,
\eqno{(\arabic{equation}c)}
$$
$$
 {\bf T}_k = -t\left[ \begin{array}{ccccc}
  M_1        & e^{-ik_y} & 0      & 0          & e^{ik_y}  \\
  e^{ik_y}   & M_2       & \ddots & 0          & 0          \\
  0          & \ddots    & \ddots & \ddots     & 0          \\
  0          & 0         & \ddots & M_{q-1}    & e^{-ik_y} \\
  e^{-ik_y}  & 0          & 0     & e^{ik_y}   & M_q
 \end{array} \right] ~,
\eqno{(\arabic{equation}d)}
$$
and
$$
 {\bf V}_\sigma = \left[ \begin{array}{ccccc}
 -\frac{\sigma mU}{2} & 0     & 0      & 0     & 0     \\
 0     & -\frac{\sigma mU}{2} & 0      & 0     & 0     \\
 0     & 0     & \ddots & 0     & 0     \\
 0     & 0     & 0      & -\frac{\sigma mU}{2} & 0     \\
 0     & 0     & 0      & 0     & -\frac{\sigma mU}{2}
 \end{array} \right]  
\eqno{(\arabic{equation}e)}
$$
with $M_n = 2\cos(k_x+ng)$.
The summation $\sum'_{{\bf k}}$ is
 over the reduced Brillouin zone,
 $\{(k_x, k_y) | -\frac{\pi}{q}\le k_x \le\frac{\pi}{q},
 ~~ -\frac{\pi}{2} \le k_y \le \frac{\pi}{2} \}$.
The Brillouin zone is reduced by $1/q$
 because there exist $q$ plaquettes per magnetic unit cell,
 and is further reduced by $1/2$
 as a result of staggered magnetization (or antiferromagnetic order).
The diagonal matrix ${\bf T}_k$ is associated with electron hopping
 and contains information on the phase modulation 
 of hopping electrons under the influence of the external field.
The off-diagonal matrix ${\bf V_\sigma}$ represents
 the antiferromagnetic electron correlation.

From the eigenvalue equation of the Hamiltonian matrix ${\bf H}_{k\sigma}$
 with the identity matrix ${\bf I}$,
\begin{equation}
 \det({\bf H}_{k\sigma} -E_k {\bf I}) = 0 ~~,
\label{genhar}
\end{equation}
 we obtain the quasiparticle energy dispersion $E_k$ of
 antiferromagnetically correlated electrons
 in the presence of magnetic field.
In the limiting case of noninteracting electrons ($U=0$)
 the `generalized' Harper's equation (\ref{genhar}) above
 is reduced to the original Harper's equation
 derived by Hasegawa et al.\cite{hasegawa}, that is,
\begin{equation}
 \det({\bf T}_{k} - \varepsilon_k {\bf I}) = 0 ~~,
 \label{harper}
\end{equation}
 where $\varepsilon_k$ is the energy dispersion of
 noninteracting electrons in the presence of magnetic field.
Following Hasegawa et al.\cite{hasegawa},
 Eq. (\ref{harper}) can be rewritten in a further simplified form,
\begin{equation}
 \gamma (\varepsilon) = \cos(qk_x) +\cos(qk_y) ~~,
\end{equation}
 where $\gamma(\varepsilon)$ is given in Table \ref{harpertable}~
 for various values of $p/q$
 including Hasegawa et al.'s results.\cite{hasegawa}
Now for the case of antiferromagnetically correlated electrons
 we obtain from the diagonalization of
 the Hamiltonian matrix ${\bf H}_{k\sigma}$ in Eq. (\arabic{Hks}c) above,
\begin{equation}
 E_k = \sqrt{\varepsilon_k + \Delta^2} ~~,
 \label{Ek}
\end{equation}
 with the band gap, $2\Delta=mU$.
The band gap $2\Delta$ is seen to depend on
 the magnitude of both the staggered magnetization $m$ and
 the electron correlation strength (Coulomb repulsion) $U$.

\begin{figure}[hbt]
\centering
\vspace*{1cm}
\epsfig{file=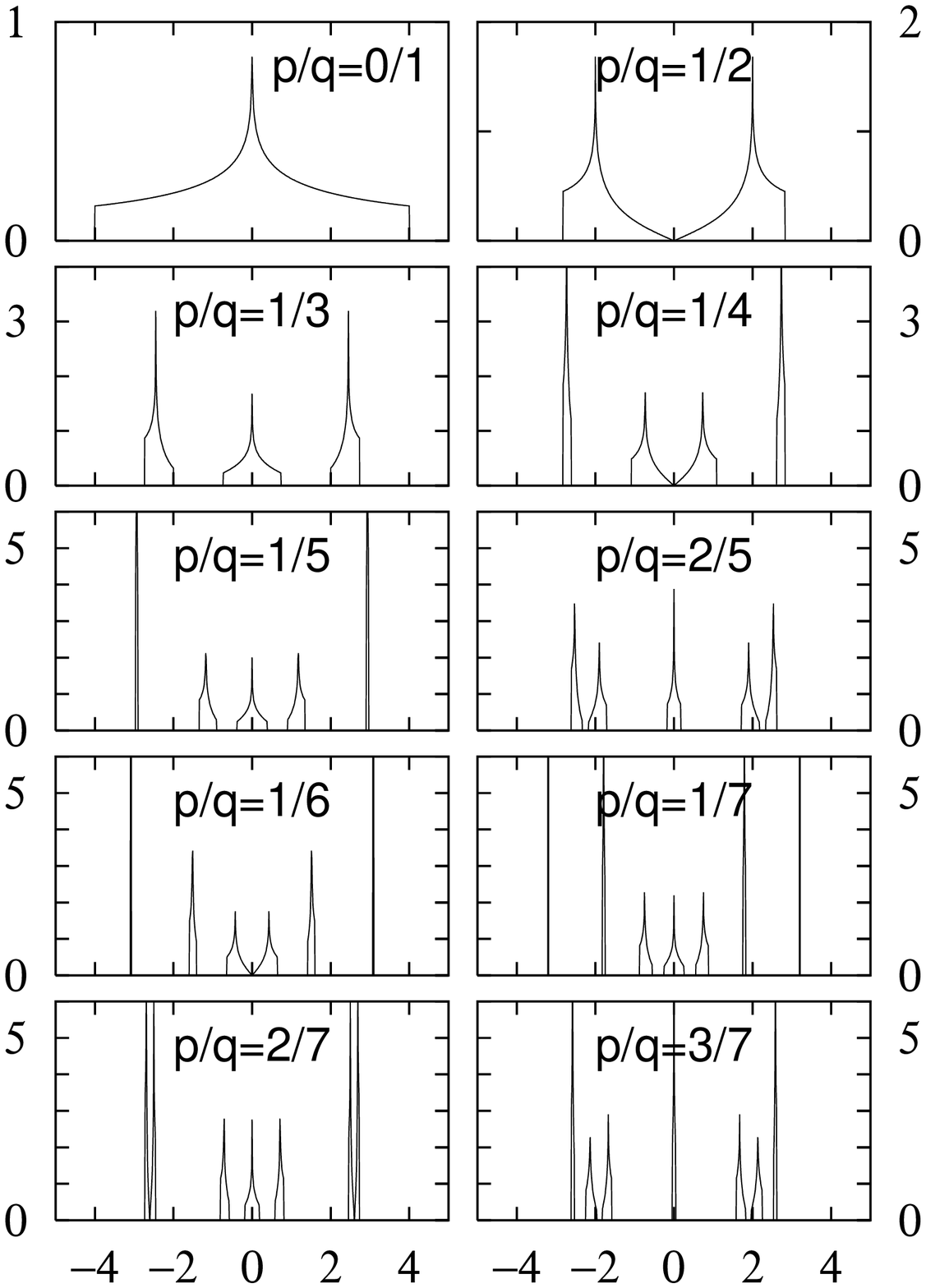, width=14cm} \\[5mm]
\caption{}
\label{dos}
\end{figure}

The energy dispersion relation Eq. (\ref{Ek}) above leads to
 the density of states,
\begin{eqnarray}
 g(E) &=& 4 \int' \frac{d^2 k}{(2\pi)^2} \delta(E-E_k)
 \nonumber \\
 &=& \frac{2}{q\pi^2} \left| \frac{d\gamma(\varepsilon)}{d\varepsilon} \right|
 K\left( \sqrt{1-\left(\frac{\gamma(\varepsilon)}{2}\right)^2} \right)
 \left| \frac{E}{\varepsilon} \right| ~~,
 \label{gE}
\end{eqnarray}
where $|\varepsilon| = \sqrt{E^2 - \left(\frac{mU}{2}\right)^2}$
and $K$ is the complete elliptic integral of the first kind,\cite{economou}
\begin{eqnarray*}
 K(\alpha) = 
 \int_0^{\frac{\pi}{2}} \frac{d\phi}{\sqrt{1-\alpha^2\sin^2\phi}} ~~.
\end{eqnarray*}
Eq. (\ref{gE}) above represents
 the density of states 
 for the systems of correlated electrons
 in the presence of magnetic field.
In the limiting case of vanishing electron correlation, i.e., $U=0$,
 it becomes exact.
In Fig.~\ref{dos} we display 
 the density of states with $U=0$
 predicted from the analytic expression of 
 the density of states in Eq. (\ref{gE}).
Encouragingly the analytic result is
 in excellent agreement with 
 the numerical results obtained by Hasegawa et al.\cite{hasegawa}
In the presence of magnetic field with the flux quanta per plaquette $p/q$,
 a single band splits into $q$ subbands.
This is analogous to the energy level splitting into
 Landau levels for electrons
 embedded in a continuum state under a magnetic field.

\section{Phase Diagram and Reentrant Behavior of Staggered Magnetization
 In Applied Magnetic Fields}
The staggered magnetization $m$ and the chemical potential $\mu$
 vary with both the temperature $T$ and the doping rate $\delta$;
 they are obtained from 
 the following self-consistent mean field equations,
 which are derived from the use of Eqs. (\arabic{uniform}),
\addtocounter{equation}{1}
\newcounter{self}
\setcounter{self}{\arabic{equation}}
$$
 1 = \int dE \frac{g(E)}{4} \frac{U}{2E}
       \left[ \tanh \frac{E^{\scriptscriptstyle(+)}}{2T} 
       - \tanh \frac{E^{\scriptscriptstyle(-)}}{2T} \right] ~~,
   \eqno{(\arabic{equation}a)} \\
$$
$$
 \delta = \int dE \frac{g(E)}{4} 
       \left[ \tanh \frac{E^{\scriptscriptstyle(+)}}{2T} 
       + \tanh \frac{E^{\scriptscriptstyle(-)}}{2T} \right] ~~,
   \eqno{(\arabic{equation}b)} 
$$
where $E^{\scriptscriptstyle (\pm)} = \pm E + \frac{U}{2} (1-\delta) - \mu$.

\begin{figure}[tbh]
\centering
\epsfig{file=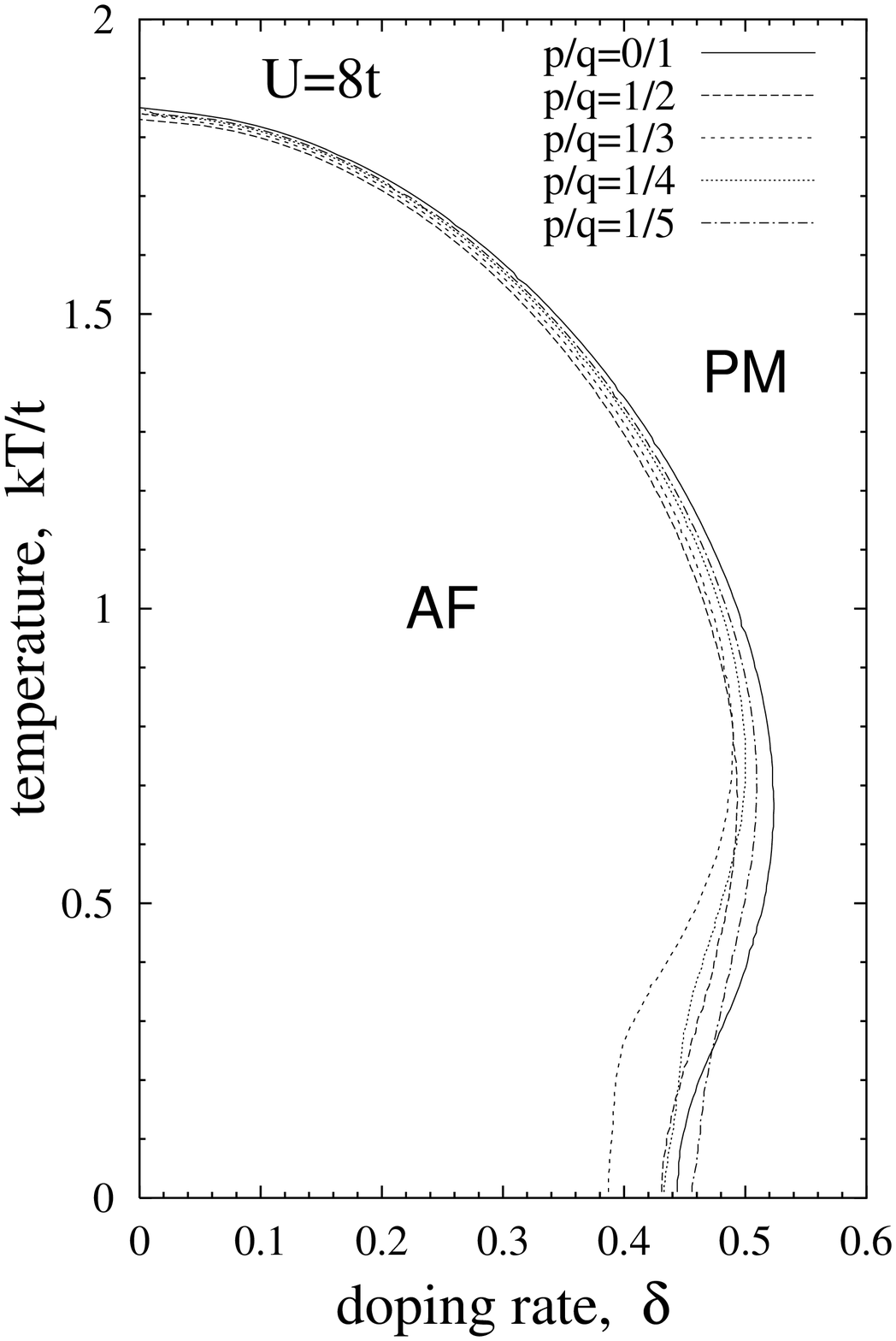, width=8cm} 
\epsfig{file=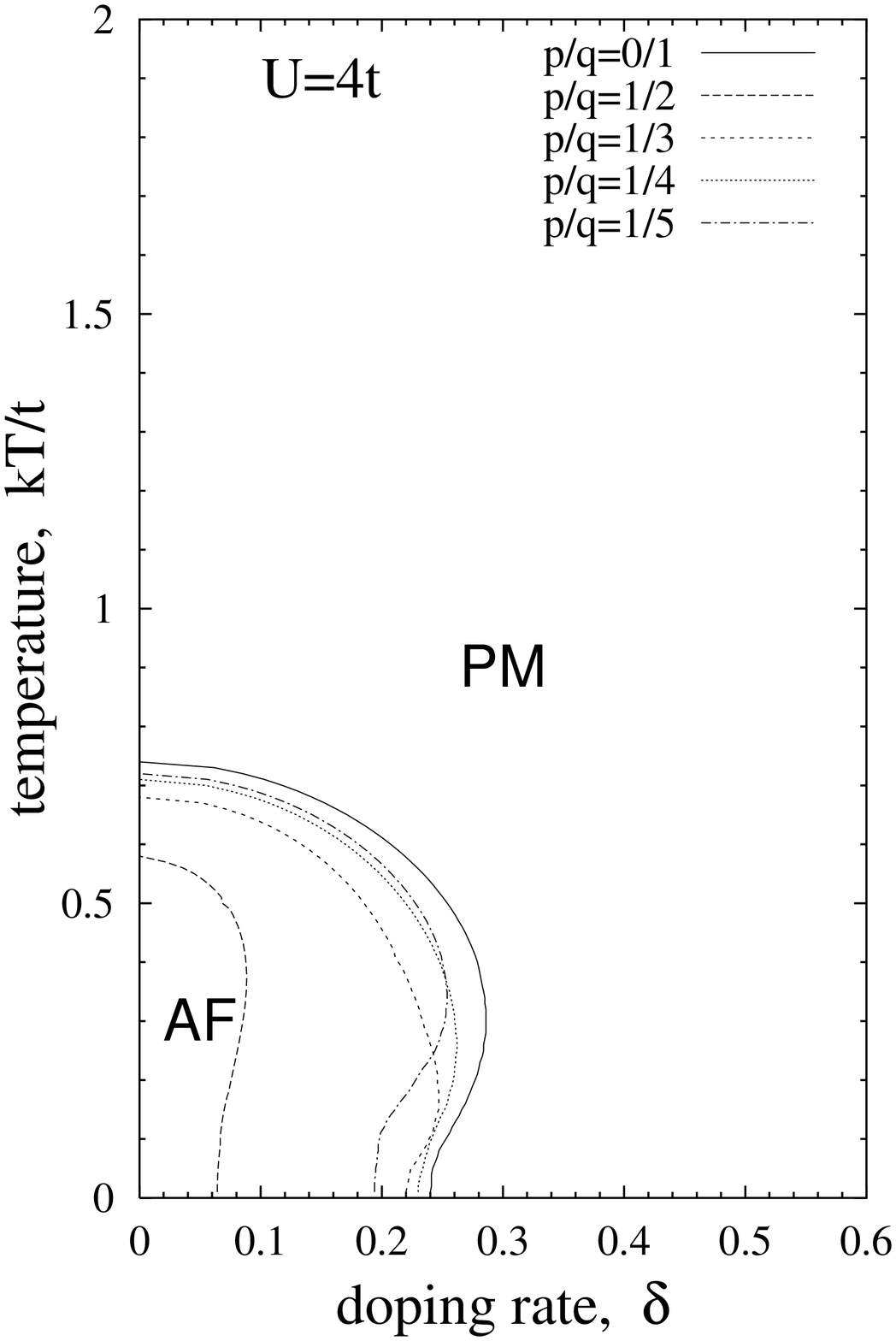, width=8cm} \\
(a) \hspace*{7cm} (b) \\[10mm]
\caption{}
\label{boundary}
\end{figure}

We examine the dependence of 
 staggered magnetization on the external magnetic field 
 in the $T$-$\delta$ plane,\cite{hong}
 by numerically solving Eqs. (\arabic{self})
 for $m$ and $\mu$ at each temperature $T$ and doping rate $\delta$.
The phase diagram of the staggered magnetization 
 in the $T$-$\delta$ plane is displayed 
 in Fig.~\ref{boundary} for several values of $p/q$.
Interestingly we find that the reentrant behavior 
 of the staggered magnetization appears 
 even in the presence of the external magnetic field.
However, the reentrant behavior of staggered magnetization 
 in the absence of magnetic field
 has earlier been discovered
 by other investigators.\cite{halvorsen,inaba,hafu}
Halvorsen et al.\cite{halvorsen} found the reentrant behavior
 using the Hubbard model 
 within the self-consistent second-order
 weak $U$-perturbation treatment.
Their studies are limited to a narrow range with small $U$
 compared to our present approach
 which can deal with the entire range of $U$.
Recently Inaba et al.\cite{inaba} found 
 a similar reentrant behavior
 using the $t$-$J$ Hamiltonian in the slave-boson representation.
However their studies refer to the case of a large $U$ limit
due to the use of the $t$-$J$ Hamiltonian.
Unlike our present study 
 their studies above refer to 
 the reentrant behavior in the absence of the external magnetic field.

\begin{figure}[tbh]
\centering
\epsfig{file=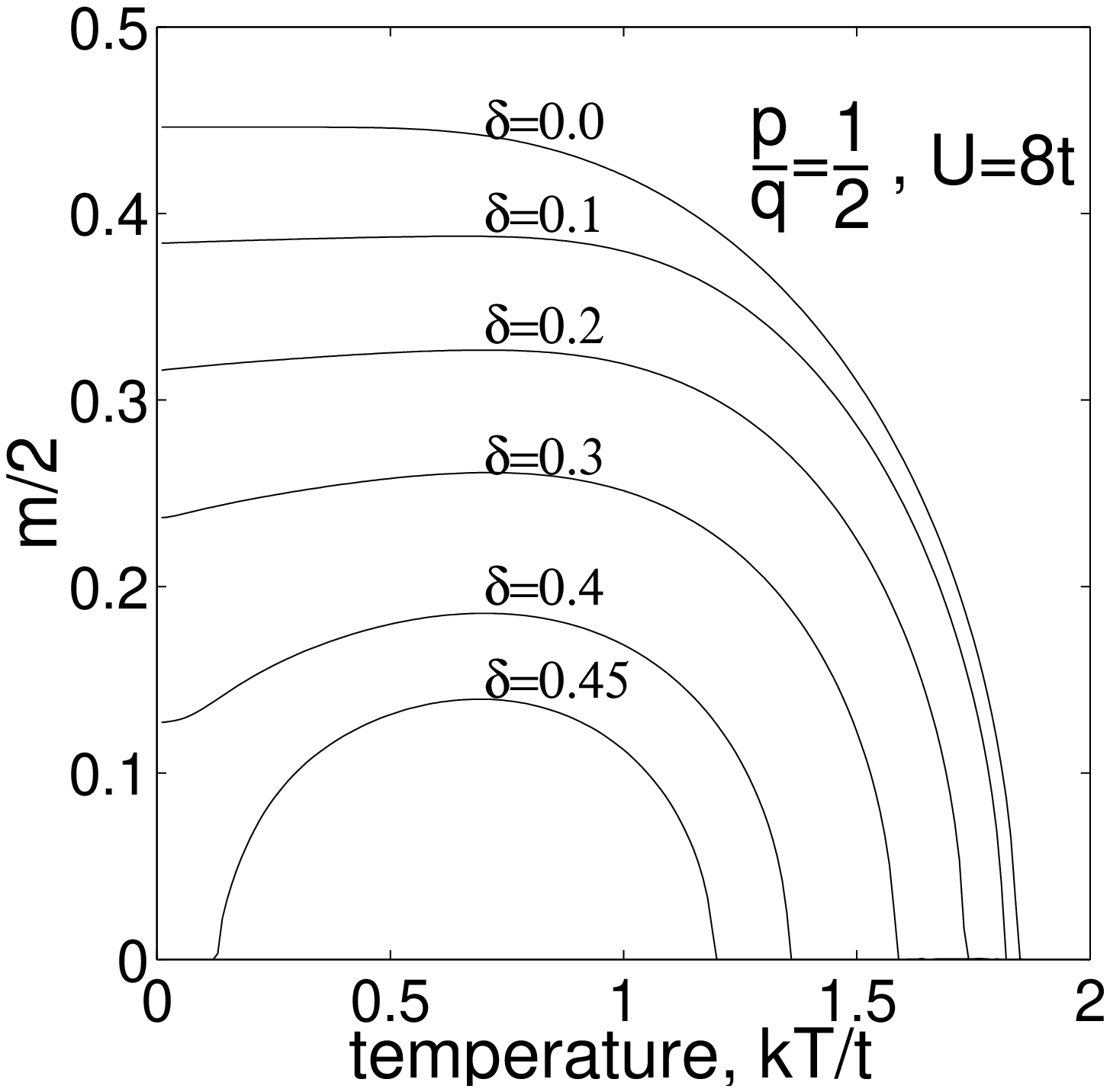, width=10cm} \\[5mm]
\caption{}
\label{reentrance}
\end{figure}

\begin{figure}[h!]
\centering
\epsfig{file=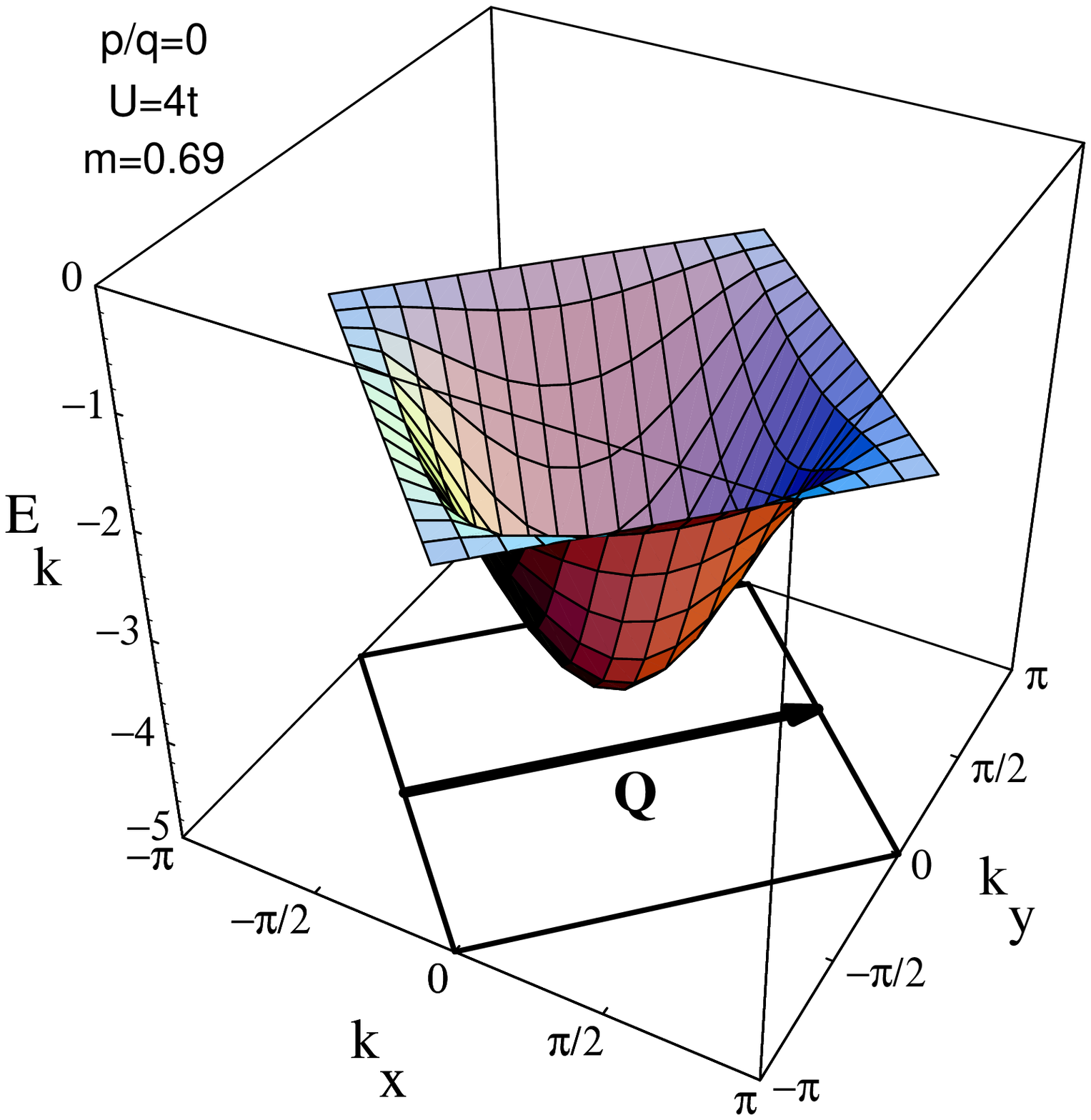, width=8cm}
\epsfig{file=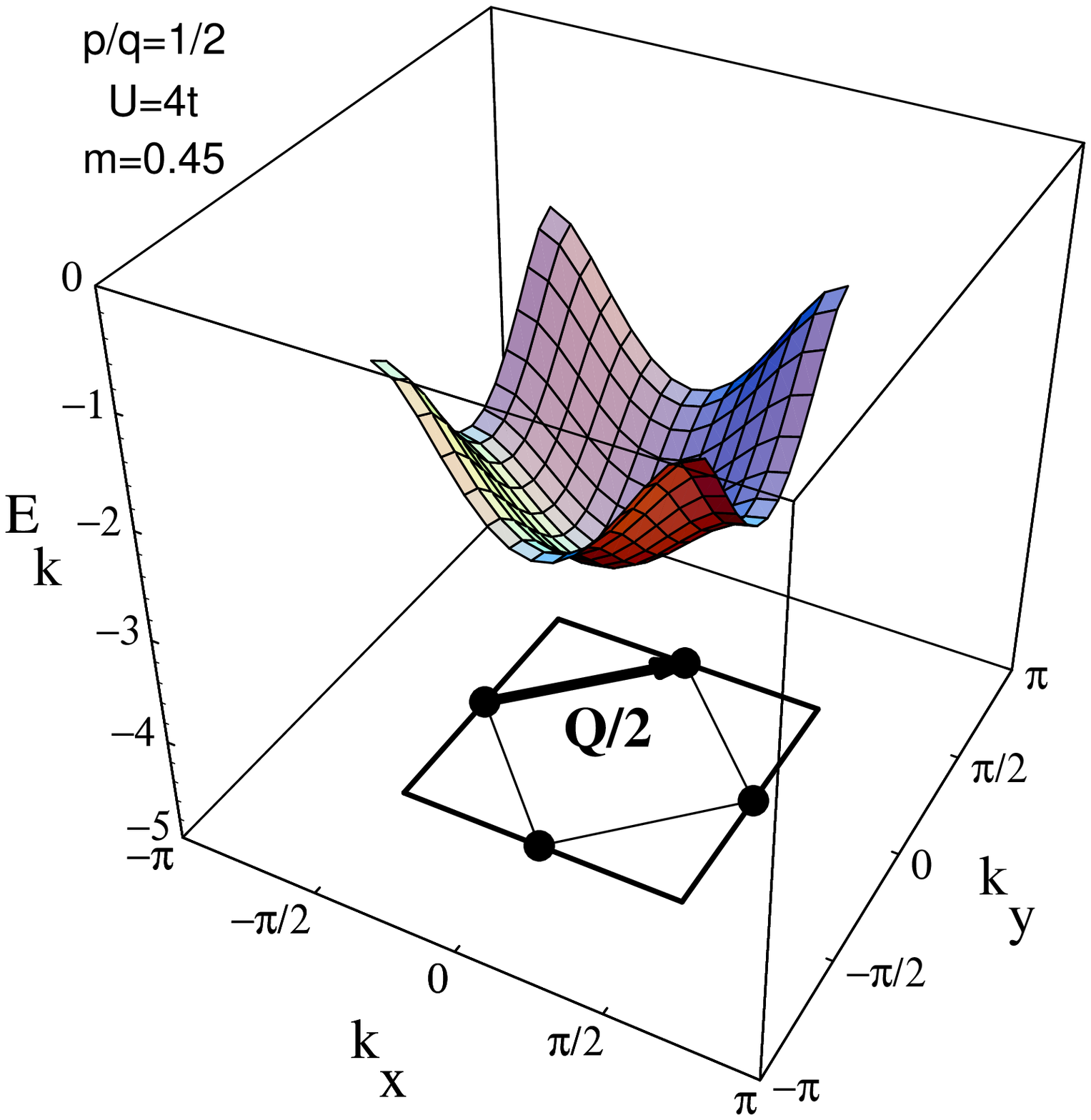, width=8cm} \\
(a) \hspace*{8cm} (b) \\[10mm]
\epsfig{file=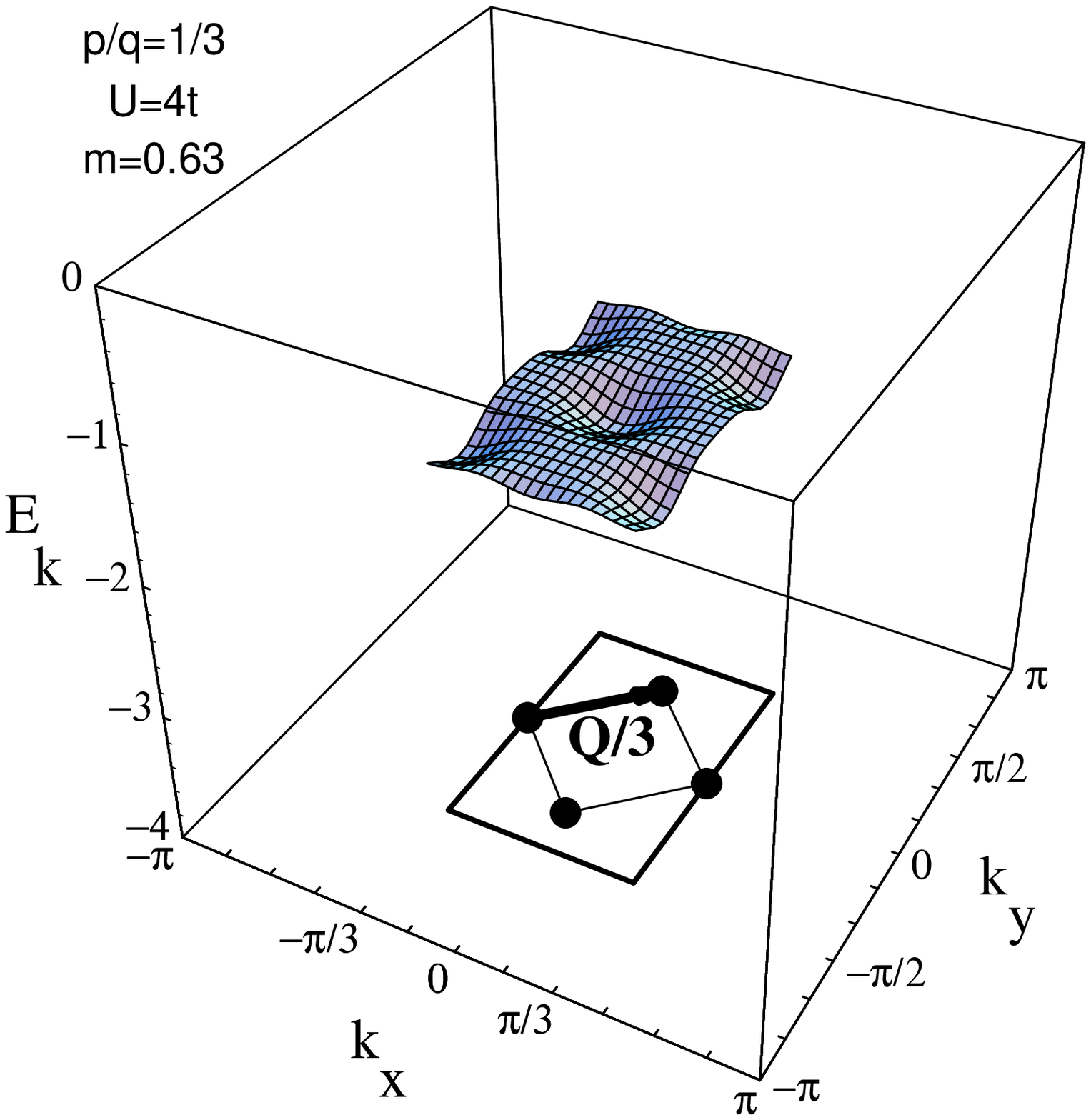, width=8cm}
\epsfig{file=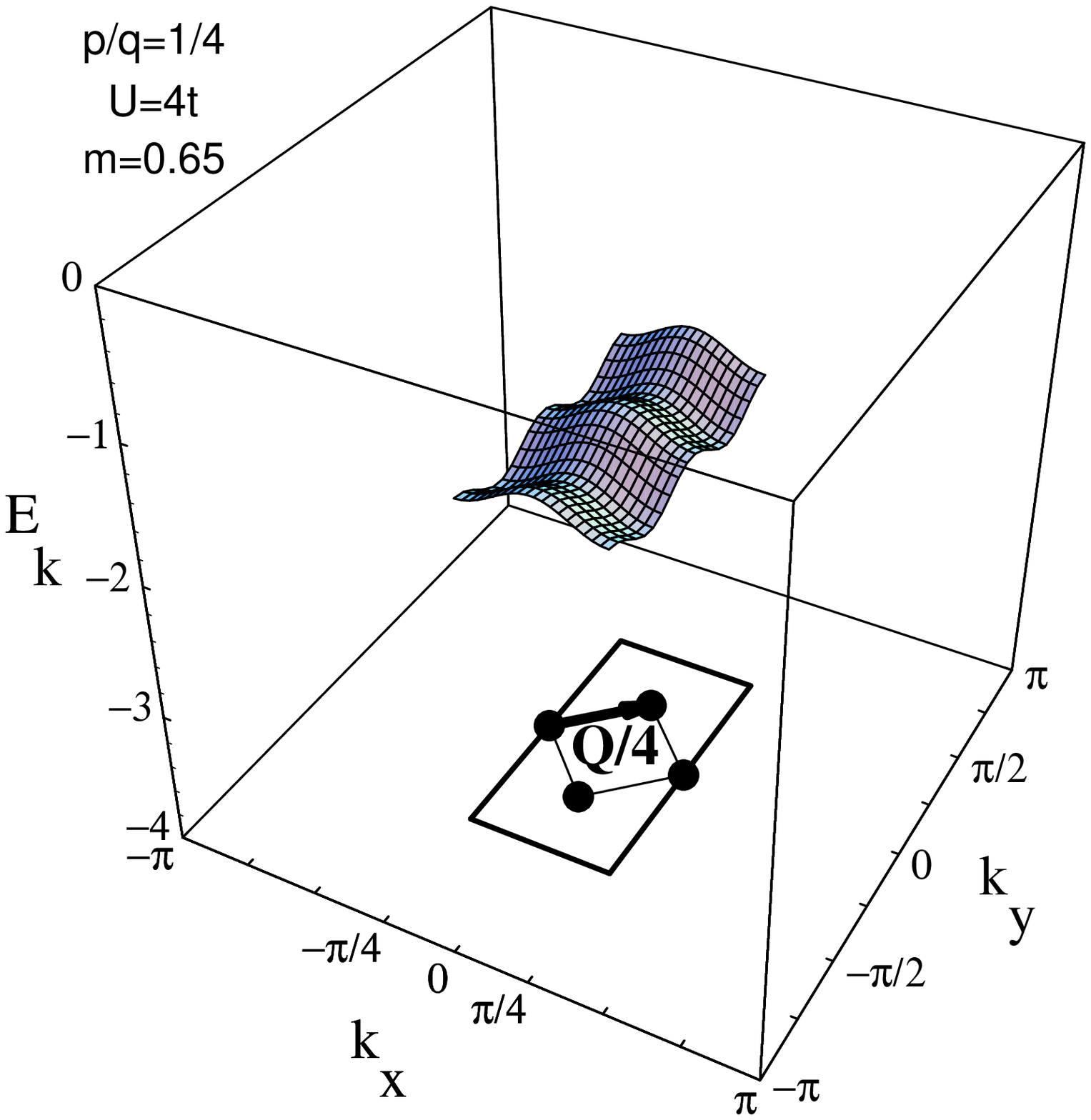, width=8cm} \\
(c) \hspace*{8cm} (d) \\[10mm]
\caption{}
\label{disurf}
\end{figure}

\begin{figure}[h!]
\centering
\epsfig{file=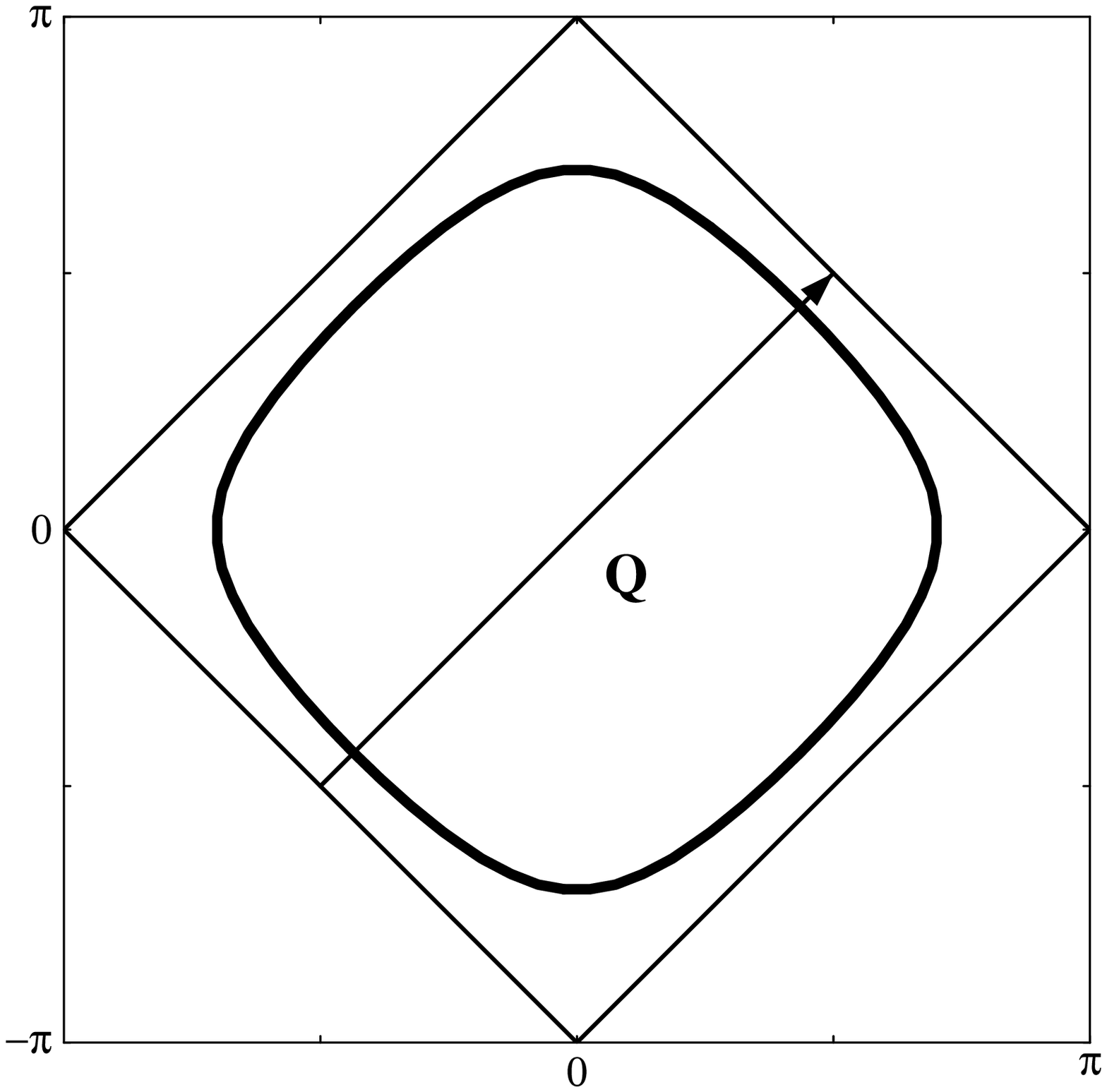, width=8cm}
\epsfig{file=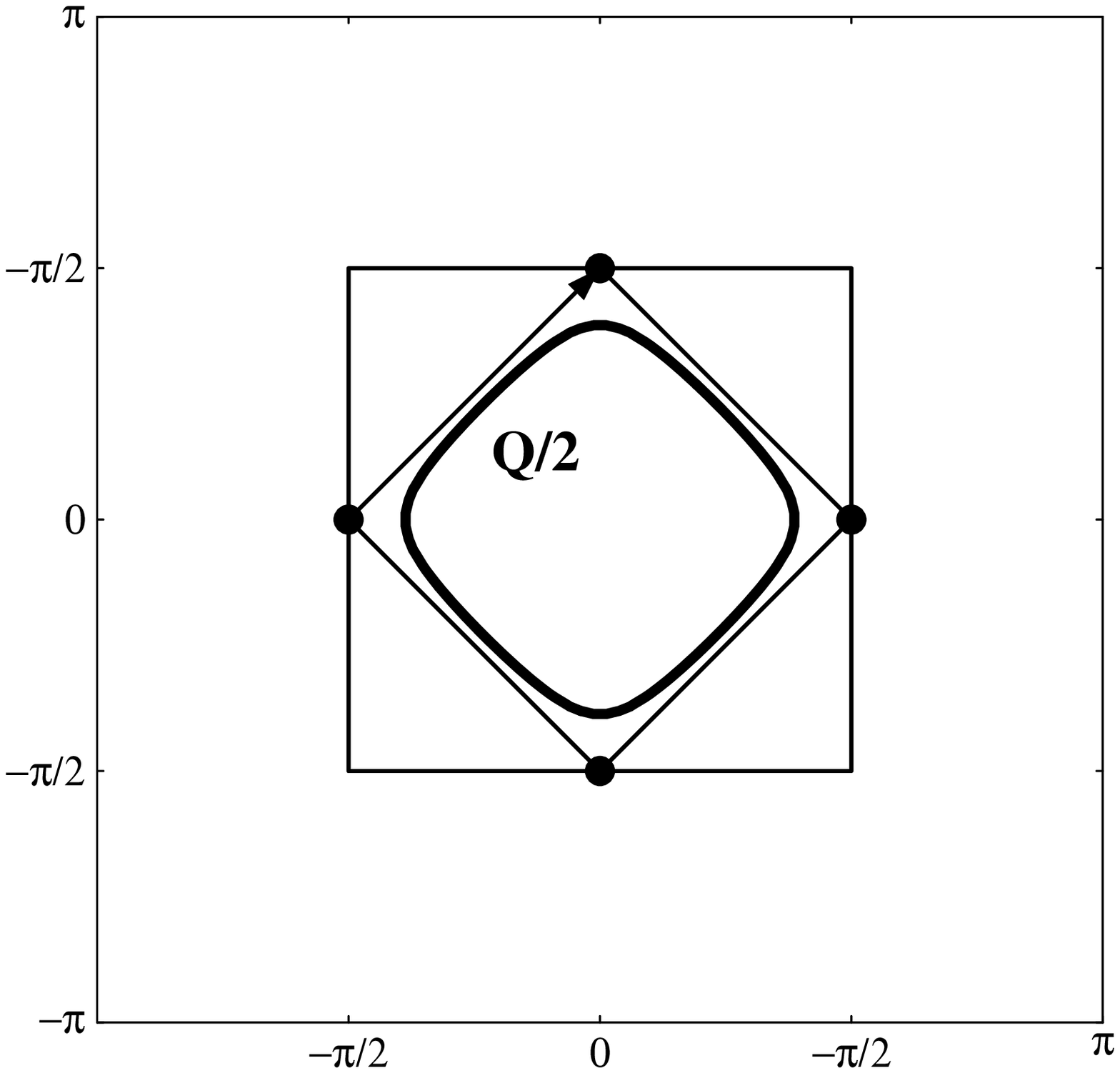, width=8cm} \\
(a) \hspace*{8cm} (b) \\[10mm]
\epsfig{file=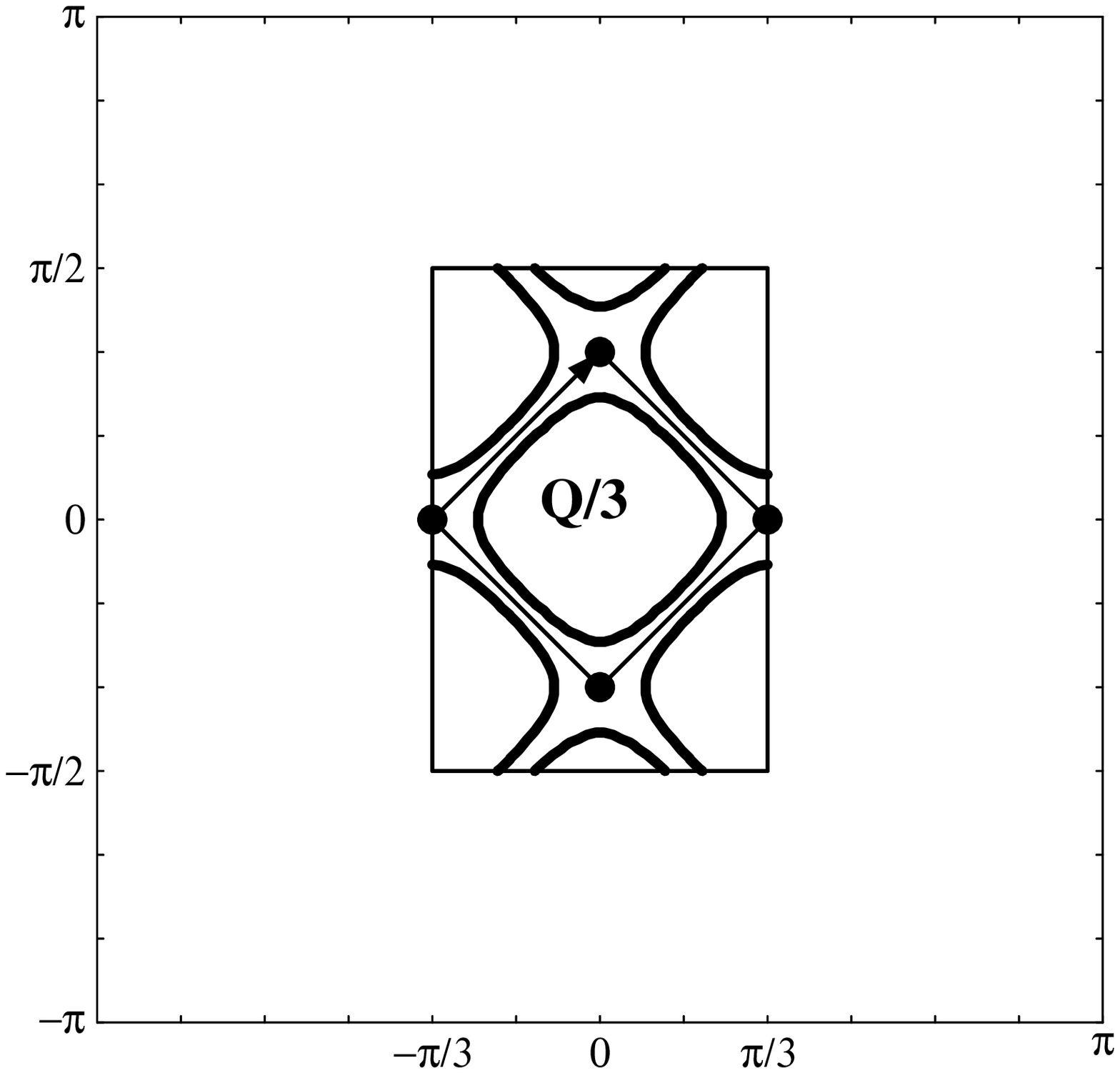, width=8cm}
\epsfig{file=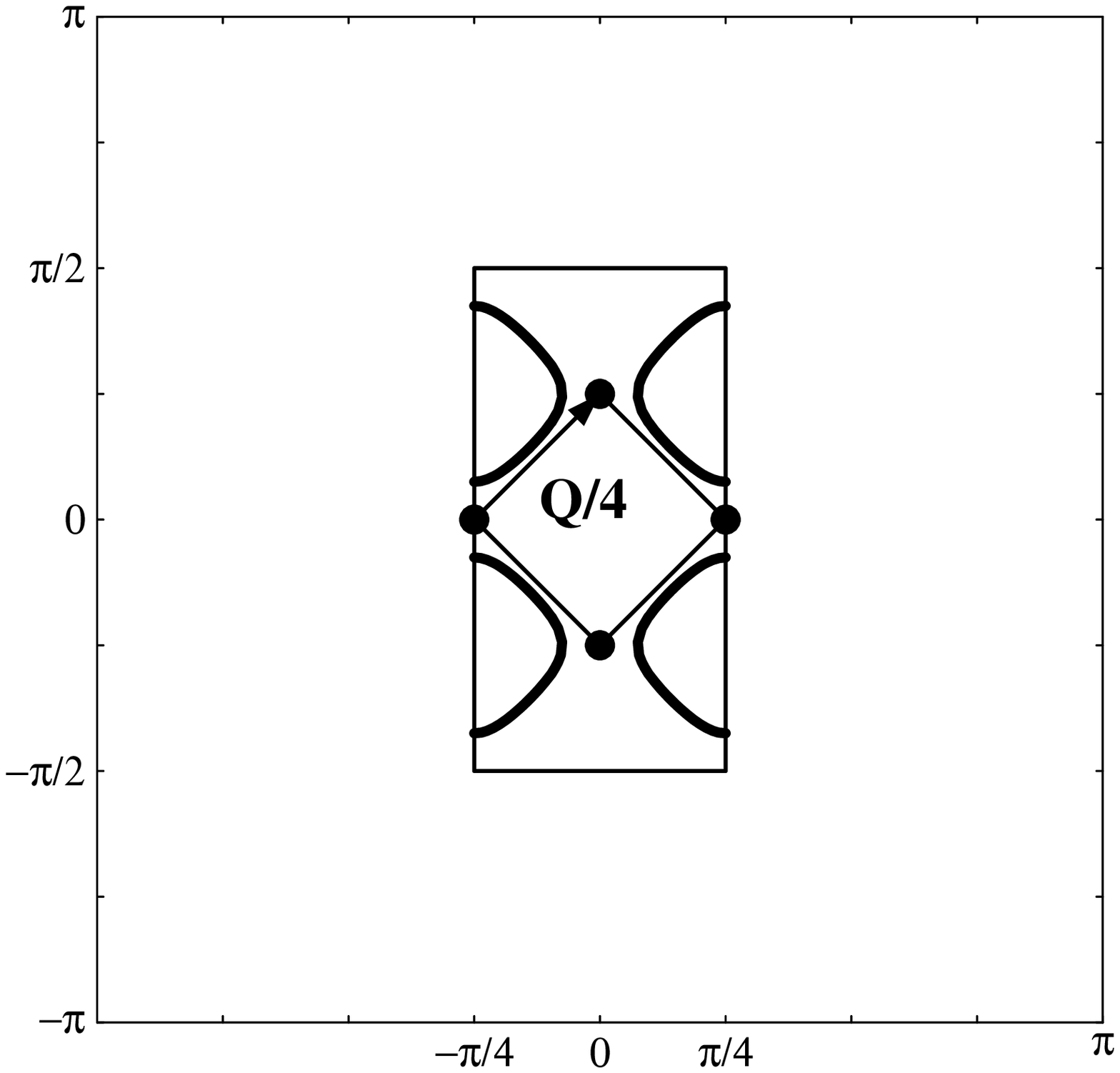, width=8cm} \\
(c) \hspace*{8cm} (d) \\[10mm]
\caption{}
\label{fermi}
\end{figure}

We now investigate
 the reentrant behavior of the staggered magnetization in detail.
In Fig.~\ref{reentrance} we display the staggered magnetization
 $m$ at $p/q=1/2$ as a function of temperature $T$
 at various doping rates $\delta$.
At half filling ($\delta=0$)
 the staggered magnetization reaches a maximum value at zero temperature.
On the other hand, away from half filling ($\delta \neq 0$)
 the predicted staggered magnetization shows a maximum
 at a finite temperature.
Above this temperature the reentrant behavior of
 a paramagnetic state is predicted.
The cause of the reentrant behavior 
 can be explained from the nesting property
 of the energy surface at saddle points.\cite{inaba,hong}
In Fig.~\ref{disurf} we display
 the variation of energy dispersion 
 of the highest occupied subband
 with the external magnetic field (or $p/q$).
The saddle points in the Brillouin zone are denoted by
 black dots at the bottom of the graph.
The adjacent saddle points are separated 
 by the well-defined nesting vector of ${\bf Q}/q$,
 by which the staggered magnetization is defined.
In Fig.~\ref{fermi}
 we display the variation of Fermi surface (thick solid lines)
 with the magnetic field 
 at finite doping rates.
As the temperature increases,
 the Fermi surface tends to smear out,
 which causes increment in the number of nesting channels,
 and consequently the staggered magnetization arises.
As the temperature still increases,
 further smearing of the Fermi surface opens
 other channels than the nesting channels.
As a result the staggered magnetization 
 will eventually disappear to allow transition 
 to a paramagnetic state.
This feature is well depicted in Figs. \ref{boundary} and \ref{reentrance}.

\section{Staggered Magnetization at Zero temperature and at Half Filling
 In a Magnetic Field}
\begin{figure}[h!]
\centering
\epsfig{file=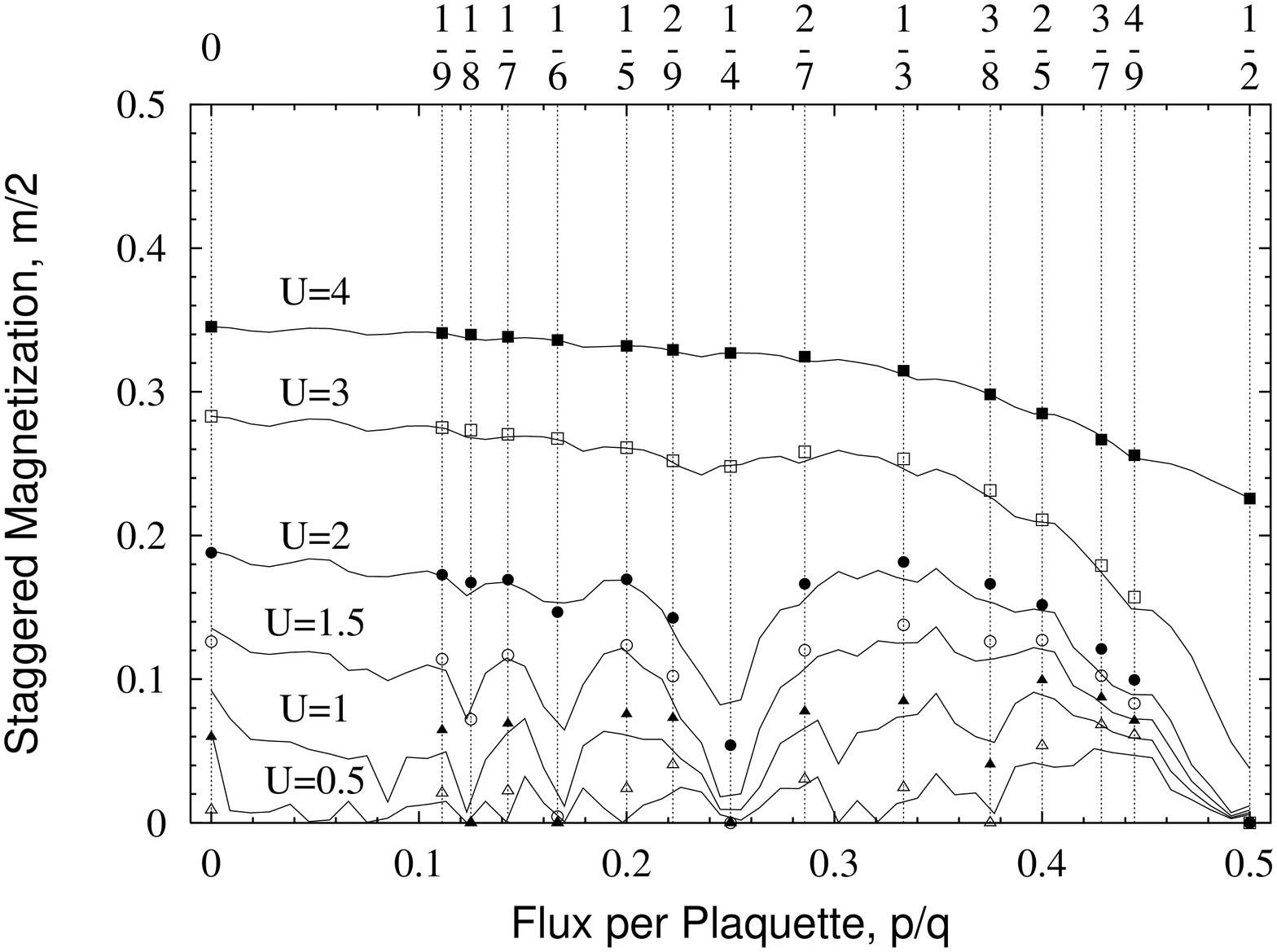, width=10cm, angle=-90}
\caption{}
\label{undulatory}
\end{figure}

Now we investigate the staggered magnetization
 at half filling $\delta=0$ and at $T=0$K
 in a magnetic field.
The chemical potential is given by $\mu=\frac{U}{2}$,
 and Eq. (\arabic{self}b)
 is trivially identified since the right hand side of the equation
 becomes zero, thus satisfying the condition of half filling,
 that is, $\delta=0$ as it should be.
Eq. (\arabic{self}a) can be expressed as
\begin{equation}
  \frac{1}{U} =
  \int_{-\infty}^0 d\varepsilon \frac{g_0(\varepsilon)}{2} 
  \frac{1}{\sqrt{\varepsilon^2+\left(\frac{mU}{2}\right)^2}}
  ~~.
 \label{Uc}
\end{equation}
Here $g_0(\varepsilon)$ is the density of states 
 of noninteracting electrons in the presence of magnetic field.
It is easily obtained from Eq. (\ref{gE}).
Eq. (\ref{Uc}) is in a similar form to the gap equation 
 that appears in the spin density wave theory 
 of cuprate materials.\cite{schrieffer,dagotto94}
 We will use it for the determination of 
 the staggered magnetization $m$
 for the system of interacting electrons
 (with correlation strength $U$)
 in the presence of magnetic field.
In Fig.~\ref{undulatory}
 the oscillatory staggered magnetization is displayed
 as a function of magnetic field (specifically $p/q$),
 for several chosen values of correlation strength $U$.
The solid lines are the results of
 self-consistent calculations for
 a $20\times 20$ finite square lattice.\cite{doh}
Various other symbols represent the results
 from the newly derived analytic relation (\ref{Uc}) above.
Encouragingly they are in good agreement
 with the self-consistent calculations 
 for the finite size square lattice.

In the following
 we explain the oscillatory behavior of staggered magnetization
 observed in our earlier work.\cite{doh}
At even denominator values of $q$ in $p/q$
 the staggered magnetization
 is predicted to disappear
 (e.g., see the case of $p/q=1/2$).
This feature is well depicted in Fig. \ref{undulatory}.
We now define the critical electron correlation strength
 (Coulomb repulsion) $U_{p/q}$
 as a value below which
 the staggered magnetization vanishes, i.e., $m=0$.
The predicted staggered magnetization
 from Eq. (\ref{Uc}) vanishes
 at the even $q$ values
 below a critical value $U_{p/q}$, i.e., $U<U_{p/q}$.
On the other hand
 in the absence of magnetic field
 the staggered magnetization tends to appear
 even at small values of $U$.\cite{dagotto94}
The critical correlation strength $U_{p/q}$ is obtained by substituting $m=0$
 in Eq. (\ref{Uc}),
 and is shown for various values of $p/q$ in Table \ref{Uctable}.
For odd $q$ in $p/q$,
 the integral in Eq. (\ref{Uc}) is logarithmically divergent,
 and thus $U_{p/q}$ does not exist.
Although not numerically precise
 for the case of finite size calculations (solid lines),\cite{doh}
 with the even denominator values of $q$ in $p/q$
 a propensity of vanishing staggered magnetization
 is seen below $U_{p/q}$
 as shown in Fig.~\ref{undulatory}.
Thus the oscillatory behavior of the staggered magnetization
 is found to occur owing to its disappearance distinctively 
at the even denominator values of $q$ in $p/q$.

\section{Conclusion}
We derived a generalized Harper's equation
 for the energy dispersion relation of
 interacting (antiferromagnetically correlated) electrons
 in an external magnetic field.
Unlike the original Harper's equation
 which deals only with noninteracting electron systems,
 the generalized Harper's equation
 derived in Eq. (\ref{genhar}) (with Eq. (\arabic{Hks}c))
 has a definite merit of studying the physical properties of 
 correlated electron systems 
 in the presence of the external magnetic field.
From this Harper's equation
 we derived an analytic formula
 for the density of states of the antiferromagnetically correlated electrons
 in the magnetic field.
For the limiting case of noninteracting electrons,
 the analytic equation for the density of states is found to be
 in good agreement with the numerical work of Hasegawa et al.\cite{hasegawa}
Further we presented the phase diagram of staggered magnetization
 in the plane of temperature $T$ vs doping rate $\delta$ 
 as a function of magnetic field
 (specifically, flux quanta per plaquette $p/q$)
 and correlation strength $U$.
From this study we demonstrated 
 a possibility of reentrant behavior of staggered magnetization
 even in the presence of the applied magnetic field.
A more accurate account of electron correlations
 beyond the mean field approximations may not 
 alter the qualitative nature of the reentrant behavior
 that we discovered in this study.
Although not reported here,
 we find from the exact diagonalization study of
 Hubbard model that
 the accurate account of correlations does not affect
 the qualitative finding here.
In the present study
 we neglected the Zeeman coupling.
At such large Coulomb repulsion energies as $U=8t$
 which we used in our calculations,
 we find that the Zeeman effect does not substantially alter
 the observed staggered magnetization.
For the case of half filling at zero temperature,
 we obtained a gap equation (\ref{Uc})
 for the determination of staggered magnetization
 for antiferromagnetically correlated electron systems
 at a given correlation strength $U$
 in the presence of magnetic field.
From this derivation we were able to determine
 a critical correlation strength $U_{p/q}$ 
 below which staggered magnetization disappears
 at even denominator values of $q$
 (but not at odd denominator values of $q$)
 of the magnetic flux quanta per plaquette, $p/q$.

\acknowledgments{
One (S.H.S.S) of the authors is grateful to the SRC program (1996) of
the Center for Molecular Science at KAIST.
He acknowledges the financial supports of
the Korean Ministry of Education BSRI (1997)
and POSTECH/BSRI.
He is also grateful to Professor Han-Yong Choi for earlier assistance.
}

\newpage
\centerline{TABLE CAPTIONS}
\begin{itemize}
\item[TABLE I.]
 $\gamma(\varepsilon)$ for various values of 
 magnetic flux quanta per plaquette $p/q$.
 The energy dispersion of noninteracting electrons
 at a given $p/q$
 is determined from the Harper's equation,
 $\gamma(\varepsilon) = \cos(qk_x + qk_y)$.

\item[TABLE II.]
 Critical correlation strength $U_{p/q}$  as a function of 
 magnetic flux quanta per plaquette $p/q$.
\end{itemize}

\newpage
\centerline{FIGURE CAPTIONS}
\begin{itemize}
\item[FIG. 1.]
 Density of states of noninteracting electrons
 in the presence of applied magnetic field
 for various flux quanta per plaquette $p/q$
 based on the analytic expression (\ref{gE}).

\item[FIG. 2.]
 Phase diagram of staggered magnetization
 for several values of magnetic flux quanta per plaquette $p/q$.
 Each line represents a boundary between
 the antiferromagnetic (AF) phase 
 and the paramagnetic (PM) phase.

\item[FIG. 3.]
 Variation of staggered magnetization
 as a function of temperature for various doping rates
 at $p/q=1/2$ and $U=8t$.

\item[FIG. 4.]
 Variation of quasiparticle energy dispersion surface
 of the highest occupied subband with $p/q$.
 Black dots denote the saddle points of the surface.
 Rectangles at the bottom of the graph represent
 the reduced Brillouin zones,
 and arrows, the nesting vectors
 between two adjacent saddle points.

\item[FIG. 5.]
 Variation of Fermi surfaces with $p/q$
 at zero temperature 
 and at hole doping rates,
 (a) $\delta=0.33$ (b) $\delta=0.30$
 (c) $\delta=0.10$ (d) $\delta=0.17$.
 Black dots denote the saddle points;
 rectangles, the reduced Brillouin zones,
 and arrows, the nesting vectors
 between two adjacent saddle points.

\item[FIG. 6.]
 Staggered magnetization (antiferromagnetic order) at $T=0$K
 as a function of $p/q$, the magnetic flux quanta per plaquette.
 The solid lines denote the results
 from self-consistent calculations 
 on a $20\times 20$ finite square lattice,
 and various other symbols represent the results 
 calculated from the analytic expression (\ref{Uc})
 corresponding to the square lattice of infinite size.
\end{itemize}

\newpage
\begin{table}[hbt]
 $$
 \begin{array}{cc|c} 
  \hline 
  \hline 
  ~p~  &  ~q~  &  2\gamma(\varepsilon) \\ 
  \hline
  1 & 2 & 
  {{\varepsilon}^2}-4 \\ 
  1 & 3 & 
  - {{\varepsilon}^3} + 6 \varepsilon \\ 
  1 & 4 & 
  {{\varepsilon}^4} - 8 {{\varepsilon}^2} + 4 \\ 
  1 & 5 & 
  - {{\varepsilon}^5} + 10 {{\varepsilon}^3}
  +  \varepsilon \left( -15 + 10 \cos ({{2 \pi }\over 5}) \right)  \\ 
  2 & 5 & 
  - {{\varepsilon}^5} + 10 {{\varepsilon}^3}
  +  \varepsilon \left( -20 - 10 \cos ({{2 \pi }\over 5}) \right)  \\ 
  1 & 6 & 
  {{\varepsilon}^6} - 12 {{\varepsilon}^4} + 24 {{\varepsilon}^2} -4 \\
  1 & 7 & 
  - {{\varepsilon}^7} +
     14 {{\varepsilon}^5} + 
     \left( -49 + 14 \cos ({{2 \pi }\over 7}) \right) {{\varepsilon}^3} + 
     \left( 42 - 28 \cos ({{2 \pi }\over 7}) - 
        28 \cos ({{3 \pi }\over 7}) \right) \varepsilon  \\ 
  2 & 7 & 
  - {{\varepsilon}^7} +
     14 {{\varepsilon}^5} + 
     \left( -49 - 14 \cos ({{3 \pi }\over 7}) \right) {{\varepsilon}^3} + 
     \left( 28 - 28 \cos ({{2 \pi }\over 7}) + 
        56 \cos ({{3 \pi }\over 7}) \right) \varepsilon \\ 
  3 & 7 & 
  - {{\varepsilon}^7} +
     14 {{\varepsilon}^5} + 
     \left( -56 - 14 \cos ({{2 \pi }\over 7}) + 
        14 \cos ({{3 \pi }\over 7}) \right) {{\varepsilon}^3} + 
     \left( 56 + 56 \cos ({{2 \pi }\over 7}) - 
        28 \cos ({{3 \pi }\over 7}) \right) \varepsilon \\
  1 & 8 & 
  {{\varepsilon}^8} - 
    16 {{\varepsilon}^6} +
    \left( 72 - 8 {\sqrt{2}} \right)  {{\varepsilon}^4} +
    \left( -96 + 32 {\sqrt{2}} \right)  {{\varepsilon}^2} + 4 \\
  3 & 8 & 
  {{\varepsilon}^8} - 
    16 {{\varepsilon}^6} +
    \left( 72 +  8 {\sqrt{2}} \right)  {{\varepsilon}^4} +
    \left( -96 - 32 {\sqrt{2}} \right)  {{\varepsilon}^2} + 4 \\
  1 & 9 & 
  - {{\varepsilon}^9} +
     18 {{\varepsilon}^7} +
     \left( -99 + 4 {\sqrt{3}} \cos ({{\pi }\over {18}}) + 
        6 \cos ({{2 \pi }\over 9}) + 
        4 {\sqrt{3}} \cos ({{7 \pi }\over {18}}) \right) {{\varepsilon}^5}\\
  & &  + 
     \left( 186 - 
        24 {\sqrt{3}} \cos ({{\pi }\over {18}}) - 
        36 \cos ({{2 \pi }\over 9}) + 
        4 {\sqrt{3}} \cos ({{5 \pi }\over {18}}) - 
        28 {\sqrt{3}} \cos ({{7 \pi }\over {18}}) + 
        24 \cos ({{4 \pi }\over 9}) \right) {{\varepsilon}^3} \\ 
  & &  + 
     \left( -126 + 36 {\sqrt{3}} \cos ({{\pi }\over {18}}) + 
        36 \cos ({{2 \pi }\over 9}) + 
        36 {\sqrt{3}} \cos ({{7 \pi }\over {18}}) - 
        54 \cos ({{4 \pi }\over 9}) \right) \varepsilon  \\
  2 & 9 & 
  - {{\varepsilon}^9} +
     18 {{\varepsilon}^7} +
     \left( -99 + 
        4 {\sqrt{3}} \cos ({{5 \pi }\over {18}}) - 
        4 {\sqrt{3}} \cos ({{7 \pi }\over {18}}) + 
        6 \cos ({{4 \pi }\over 9}) \right) {{\varepsilon}^5}  \\
  & &  + 
     \left( 186 - 4 {\sqrt{3}} \cos ({{\pi }\over {18}}) - 
        24 \cos ({{2 \pi }\over 9}) - 
        28 {\sqrt{3}} \cos ({{5 \pi }\over {18}}) + 
        24 {\sqrt{3}} \cos ({{7 \pi }\over {18}}) - 
        60 \cos ({{4 \pi }\over 9}) \right) {{\varepsilon}^3}  \\
  & &  + 
     \left( -126 + 54 \cos ({{2 \pi }\over 9}) + 
        36 {\sqrt{3}} \cos ({{5 \pi }\over {18}}) - 
        36 {\sqrt{3}} \cos ({{7 \pi }\over {18}}) + 
        90 \cos ({{4 \pi }\over 9}) \right) \varepsilon  \\ 
  4 & 9 & 
  - {{\varepsilon}^9} +
     18 {{\varepsilon}^7} +
     \left( -99 - 4 {\sqrt{3}} \cos ({{\pi }\over {18}}) - 
        6 \cos ({{2 \pi }\over 9}) - 
        4 {\sqrt{3}} \cos ({{5 \pi }\over {18}}) - 
        6 \cos ({{4 \pi }\over 9}) \right) {{\varepsilon}^5}  \\
  & &  + 
     \left( 186 + 28 {\sqrt{3}} \cos ({{\pi }\over {18}}) + 
        60 \cos ({{2 \pi }\over 9}) + 
        24 {\sqrt{3}} \cos ({{5 \pi }\over {18}}) + 
        4 {\sqrt{3}} \cos ({{7 \pi }\over {18}}) + 
        36 \cos ({{4 \pi }\over 9}) \right) {{\varepsilon}^3}  \\ 
  & &  + 
     \left( -126 - 36 {\sqrt{3}} \cos ({{\pi }\over {18}}) - 
        90 \cos ({{2 \pi }\over 9}) - 
        36 {\sqrt{3}} \cos ({{5 \pi }\over {18}}) - 
        36 \cos ({{4 \pi }\over 9}) \right) \varepsilon  \\
  \hline 
  \hline 
 \end{array} $$ \\
 \caption{}
 \label{harpertable}
 \end{table}

\newpage
\vspace*{5cm}
\begin{table}[hbt]
 \centering
 $\begin{array}{c|ccccc}
  \hline \hline
  \frac{p}{q} & \frac{1}{8} & \frac{1}{6} & \frac{1}{4} & \frac{3}{8}
  & \frac{1}{2} \\
  \hline
  U_{p/q} & ~1.29~ & ~1.49~ & ~1.87~ & ~0.897~ & ~3.11~ \\
  \hline \hline
 \end{array}$ \\[10mm]
 \caption{}
 \label{Uctable}
\end{table}

\newpage

\newpage
\vspace*{4cm}

\newpage
\vspace*{5cm}

\newpage

\newpage

\newpage

\end{document}